\documentclass{article}

\PassOptionsToPackage{numbers, compress}{natbib}

 \usepackage{graphicx}
 \usepackage{aas_macros}
 \usepackage{wrapfig}

\usepackage[main, final]{neurips_2025}

\usepackage[utf8]{inputenc} 
\usepackage[T1]{fontenc}    
\usepackage{hyperref}       
\usepackage{url}            
\usepackage{booktabs}       
\usepackage{amsfonts}       
\usepackage{nicefrac}       
\usepackage{microtype}      
\usepackage{xcolor}         

\title{Hierarchical Simulation-Based Inference of Supernova Power Sources and their Physical Properties}

\author{%
  Edgar P. Vidal 
   \\
  Department of Physics and Astronomy\\
  Tufts University \\
  Medford, MA 02155, USA \\
  \texttt{edgar.vidal@tufts.edu} \\
  \And
  Alexander T. Gagliano \\
  Institute for AI and Fundamental Interactions \\
  Massachusetts Institute of Technology \\
  Cambridge, MA 02139, USA \\
  \texttt{gaglian2@mit.edu} \\
  \AND
    Carolina Cuesta-Lazaro \\
   Flatiron Institute \\
  Institute for Advanced Studies\\
  \texttt{ccuesta-lazaro@flatironinstitute.org} 
}

\begin{document}

\maketitle

\begin{abstract}
  Time domain surveys such as the Vera C. Rubin Observatory are projected to annually discover millions of astronomical transients. This and complementary programs demand fast, automated methods to constrain the physical properties of the most interesting objects for spectroscopic follow up. Traditional approaches to likelihood-based inference are computationally expensive and ignore the multi-component energy sources powering astrophysical phenomena. In this work, we present a hierarchical simulation-based inference model for multi-band light curves that 1) identifies the energy sources powering an event of interest, 2) infers the physical properties of each subclass, and 3) separates physical anomalies in the learned embedding space. Our architecture consists of a transformer-based light curve summarizer coupled to a flow-matching regression module and a categorical classifier for the physical components. We train and test our model on $\sim$150k synthetic light curves generated with \texttt{MOSFiT}. Our network achieves a 90\% classification accuracy at identifying energy sources, yields well-calibrated posteriors for all active components, and detects rare anomalies such as tidal disruption events (TDEs) through the learned latent space. This work demonstrates a scalable joint framework for population studies of known transients and the discovery of novel populations in the era of Rubin.
\end{abstract}
\section{Introduction}\label{sec:Introduction}

A central theme of time-domain astrophysics lies in recovering the physical properties of an astrophysical system from photometric time-series observations alone. This is particularly relevant for the explosions of stars as supernovae (SNe). While the majority of observed supernovae are powered by reprocessing of high-energy photons produced by the radioactive decay of $^{56}$Ni synthesized in the explosion, the observational taxonomy has expanded to include shock-powered SNe whose emission is dominated by interaction between the expanding ejecta and surrounding circumstellar material \citep{2017Smith_InteractingSNe} and superluminous SNe (SLSNe) powered by a compact central engine \citep{2018Moriya_SLSN}. 

While the physical signatures of these energy sources imprint themselves on SN observations, significant degeneracies exist at the level of optical photometry; worse, supernova emission can include contributions from multiple physical mechanisms simultaneously \citep{2012_Moriya}. Existing inference techniques \citep{2022Villar_SBI,2024Garretson_SBI} typically constrain the physical properties of a single energy source whose energy input is assumed to dominate during specific phases of explosion (e.g., during the photospheric or nebular phase).

A complementary objective in time-domain astrophysics is the automated detection of anomalous events. Existing approaches flag supernovae whose light curve deviates from an empirical model \citep{2022Muthukrishna_AD} or from those observed in the broader population \citep{2025Gupta_AD}. The detection of SNe whose emission deviates from expected \textit{physical} models, however, is critical to constrain the diversity of emission mechanisms and for discovering entirely novel phenomena in upcoming large-scale photometric surveys such as the Vera C. Rubin Observatory Legacy Survey of Space and Time (LSST, \citealt{2019Ivezic_LSST}). In this work, we simultaneously address both objectives with a hierarchical simulation-based inference framework capable of constraining both the primary energy sources associated with supernova emission \textit{and} the physical properties of those sources.

To generate a realistically diverse sample of physical phenomena, we generate synthetic \textit{ugrizy} LSST light curves from the Modular Open Source Fitter for Transients \citep[\texttt{MOSFiT};][]{2018Guillochon_MOSFiT}. \texttt{MOSFiT} couples semi-analytic physical models for an event's spectral energy distribution to modules that modulate this emission, as with diffusion through explosion ejecta or a viscous delay in the accretion of matter onto a black hole. We consider three energy sources in \texttt{MOSFiT}: the radioactive decay of $^{56}$Ni \citep{1994_NiCoDecay}, interaction with circumstellar material \citep[CSM;][]{2013Chatzopoulos_CSM,2017Villar_CSM,2020Jiang_CSM}, and spin-down by a central magnetar \citep{2017Nicholl_Magnetar}. We define seven total SN models consisting of all combinations of these three components, and list our priors for the associated physical parameters in Table~\ref{tbl:priors} in the Appendix.

We generate 20,000 light curves from each SN model, with 100 observations in each filter uniformly sampled in time from explosion to 200 days following explosion. We further split our dataset into fractions of 80\%/10\%/10\% for training, testing, and validation, respectively. 

\section{Hierarchical Simulation-Based Inference}\label{sec:sbi}

Our goal is to model the joint posterior distribution, $p(\mathcal{S}, \theta|x)$ over active energy source combinations $\mathcal{S}$ and their associated physical properties, parametrized as $\theta$, given the observed light curve $x$. We learn optimal summary statistics, $x_s = f_\psi(x)$, from the simulated samples, where $f_\psi$ is the shared summarizer trained jointly with the neural posterior model.

As in \cite{2023Schroder_Simultaneous}, we model the joint posterior hierarchically by learning the conditionals $p(\mathcal{S}, \theta|x_s) = p(\mathcal{S}|x_s) p(\theta|\mathcal{S},x_s)$, that we refer to as the source posterior, $p(\mathcal{S}|x_s)$, and the parameter posterior $p(\theta|\mathcal{S},x_s)$. The light curve summarizer is shared between the two components.

\paragraph{The light curve shared summarizer.}
To address the sparsity and irregular sampling of supernova light curves, we encode our \texttt{MOSFiT} light curves using the multi-band transformer model presented in \cite{2024Zhang_Maven}. Observations are provided as input in the format $\{t, m\}$ after normalization, where $t$ represents the number of days from first observation and $m$ represents the event's apparent magnitude. We project the times $t$ using sinusoidal positional encodings, concatenate all brightness measurements $m$ across all bands and add a one-hot-encoded vector representing the photometric passband of the observation (one of LSST-$ugrizy$). We use multi-head attention with 2 attention heads, and aggregate the outputs using a learnable query vector to produce a single weighted representation, $x_s \in \mathbb{R}^{d_s}$. The dimensionality of the summary statistics is set to $d_s = 64$.

\paragraph{Modeling the source posterior.} We introduce a categorical network\footnote{\cite{2023Schroder_Simultaneous} introduce a mixture of multivariate binary Grassmann distributions to predict model components in a similar hierarchical model; we find our training is more stable with the categorical network.} to model the posterior over potential supernova energy sources, $p(\mathcal{S}|x_s)$, given a light curve representation $x_s$ output by the summarizer. Let $\mathcal{C}$ be the set of source components, and let $\mathcal{S}\subseteq\mathcal{C}$ denote the composite class (an element of the power set of $\mathcal{C}$) describing all active sources. We model a categorical distribution over all non-empty subsets, i.e., over $2^{|\mathcal{C}|}-1$ classes, which allows for multiple sources components while keeping a single target per transient. Training is performed using the standard negative log likelihood loss function, $\mathcal{L}_\mathrm{source}(\psi) = -\log q_\psi(\mathcal{S}|x_s)$, where $q(\mathcal{S}|x_s)$ denotes the model distribution approximating the true posterior $p(\mathcal{S}|x_s)$, and $\psi$ are the weights of the network. The categorical network has 4 hidden layers with 128 hidden dimensions, uses sigmoid activations, and a final softmax output layer produces the predicted probabilities for each combination of source components.

\paragraph{Modeling the parameter posterior.} We model the posterior over physical parameters, $p(\theta|\mathcal{S},x_s)$, using flow matching neural posterior estimation (FMPE) \citep{2023Wildberger_FMPE}. Flow matching learns to transform samples from a base distribution, here a normal distribution, to the target posterior by predicting velocity fields. In \citep{2023Wildberger_FMPE}, the authors demonstrated that flow matching can enhance the flexibility of the more commonly used posterior estimation models, normalizing flows, whilst improving accuracy in data scarce scenarios.

In this work, we introduce a transformer as the velocity prediction network, conditioned on both the light curve summaries statistics $x_s$ and the energy source combination $\mathcal{S}$. To handle varying parameter dimensionalities across different energy sources, we mask unused parameters in the transformer. During training, we condition and mask based on the true energy sources; during inference, we use outputs from the source posterior network.
\begin{figure}[t]
    \centering
    \includegraphics[width=1\textwidth]{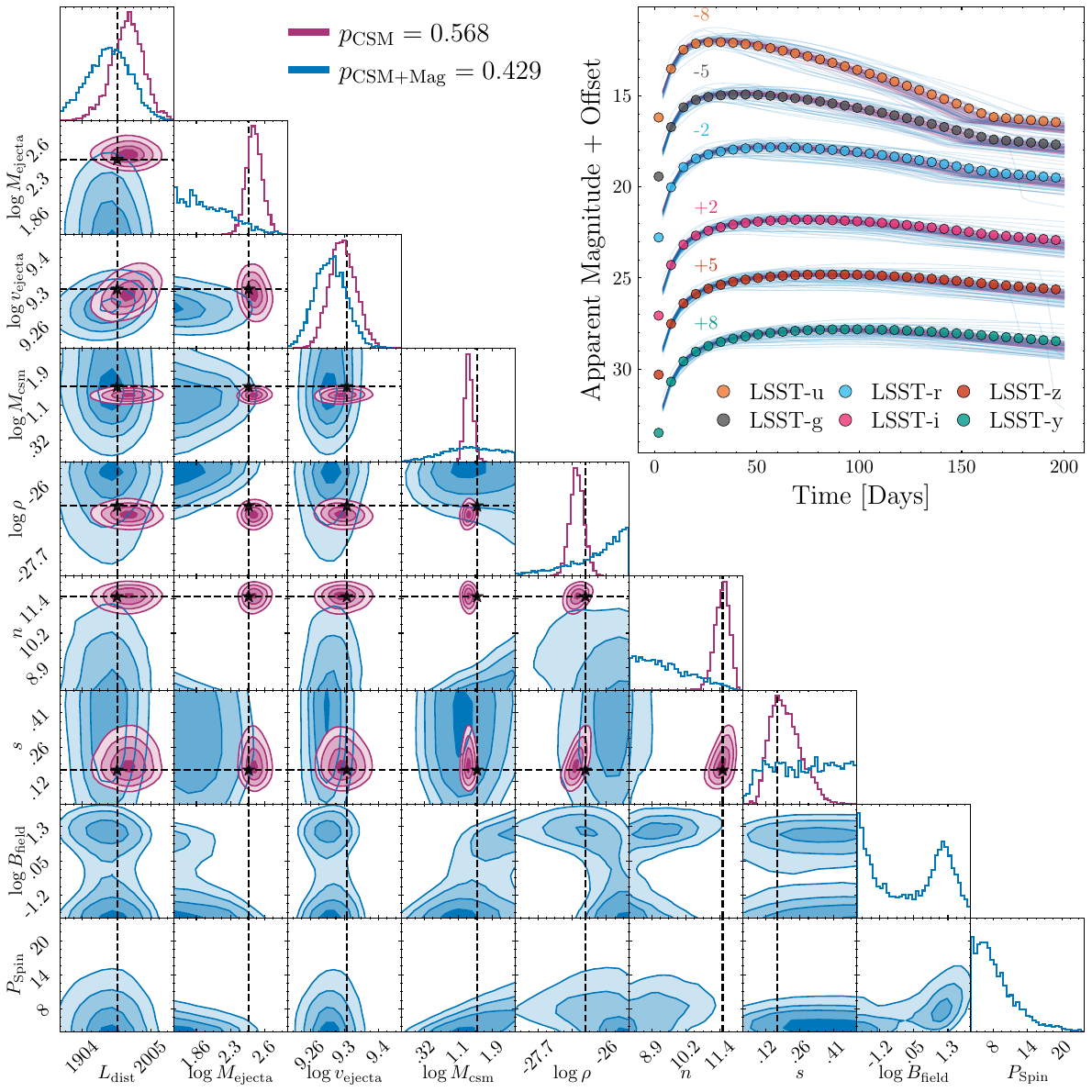}
    \caption{Example joint inference of power sources and their physical properties for a supernova light curve. On the left, we show the posterior samples for the two most likely source combinations powering the light curve, CSM (pink) and CSM+Mag (blue). We show that the presence of the magnetar engine can significantly alter the posteriors for the CSM parameters. In the upper right corner, we show the observed light curve and overlay the light curves drawn from the posteriors of the two power sources.}
    \label{fig:posteriors}
\end{figure}
We train both components jointly: $\mathcal{L} = \mathcal{L}_\theta (\phi) + \lambda \mathcal{L}_\mathrm{source}(\psi)$, where $\mathcal{L}_\theta (\phi)$ is the FMPE loss, and $\phi$ are the weights of the transformer velocity model. When $\lambda=1$, it can be shown that this loss function minimizes the expected Kullback-Leibler divergence between the true joint posterior $p(\mathcal{S}, \theta|x_s)$ and the approximate posterior $q_\phi(\mathcal{S}|x_s) q_\psi(\theta|x_s, \mathcal{S})$, under the assumption of regularity on $v_t$. Empirically, we find that $\lambda = 0.83$  gives the best approximation to the true posterior, as shown in Section~\ref{sec:results} below.

\section{Results}\label{sec:results}

\subsection{Sampling the joint posterior}
In Figure~\ref{fig:posteriors}, we show the joint inference of model components and their physical parameters for a light curve with degeneracy between energy sources predicted by the categorical network. 
 The corner plot contains posterior samples of the two source combinations with highest probability: CSM and CSM with a magnetar engine, having probabilities $p_\mathrm{CSM} = 0.568$ and $p_\mathrm{CSM+Mag} =0.439$, respectively. The presence of the magnetar engine can shift the posteriors of the CSM parameters $\rho$ and $n$ substantially, impacting the inferred properties of both the pre-explosion mass-loss history and the nature of the progenitor star (from a luminous blue variable or Wolf-Rayet-like progenitor with $7<n<10$ in the CSM+Mag model to a red supergiant-like progenitor with $n=12$ in the CSM-only model; \citealt{1999Matzner_Index}). 

In the upper right corner of Figure~\ref{fig:posteriors}, we show the light curve together with posterior resimulations with and without the magnetar engine. Whilst the CSM model fits the data well, we find that the magnetar model can also produce posterior samples that reproduce the observations. There remains active debate surrounding the dominant energy sources of superluminous supernovae; events with smooth photometric evolution and a lack of narrow spectral features have been argued to be powered by a magnetar engine \citep{2024Gomez_SLSN}, while the photometric diversity of the class may be more consistent with CSM interaction differing in e.g., geometry \citep{2025Pessi_SLSNe}. While we caution that \texttt{MOSFiT} is unable to simulate complex aspherical CSM morphologies or interaction beneath the explosion photosphere, the degeneracies in Figure~\ref{fig:posteriors} suggest that the presence of a magnetar engine may be hidden for months (as in, e.g., SN~2020wnt; \citealt{2023Tinyanont_2020wnt}), particularly when inferred from sparse and noisy light curves for events at high redshift.

We show a quantitative assessment of the learned posteriors in Appendix~\ref{ap:coverage}. Figure~\ref{fig:loss_curve} shows the results of a TARP coverage test \citep{lemos2023samplingbasedaccuracytestingposterior} demonstrating that all posteriors for all sources are well calibrated.

\subsection{Detection of physical anomalies}\label{subsec:anomalies}
\begin{wrapfigure}{r}{0.55\textwidth}
    \centering
    \includegraphics[width=0.55\textwidth]{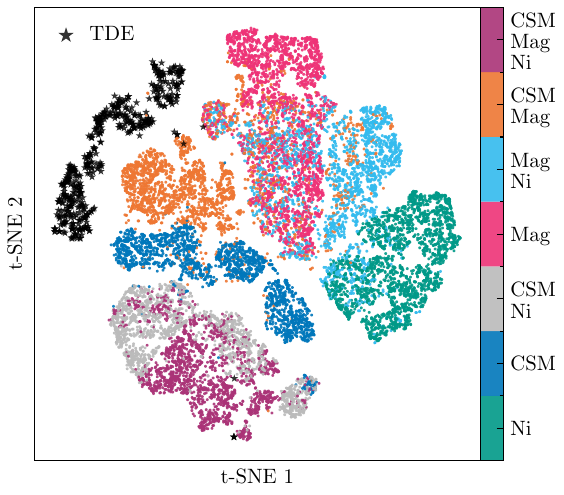}
    \caption{tSNE embedding of the latent features of all simulated SN light curves, colored by their power energy sources. TDE light curves are shown as black stars.}
    \label{fig:tsne}
\end{wrapfigure}

Next, we explore our model's capacity to identify novel time-domain phenomena. We simulate 10,000 Tidal Disruption Events (TDEs) in \texttt{MOSFiT} \citep{2019Mockler_MOSFIT}, events which are powered by the accretion of tidally-stripped stellar material onto a supermassive black hole, as our anomalous class. Our priors for the model are given in Table~\ref{tbl:priors} in the Appendix. 

We use the trained hierarchical SBI model to produce fixed-length representations of both our SN and TDE light curves, and further project them into a two-dimensional space for visualization with t-distributed Stochastic Neighbor Embedding \citep[tSNE;][]{maaten2008visualizing}. Our results are shown in Figure~\ref{fig:tsne}.

We find strong separation between individual model components and physically meaningful overlap between models with multiple energy sources. Most overlap occurs between a subset of CSM+Mag+Ni, CSM+Mag, and Mag-only light curves, suggesting that for these events the contributions from the central magnetar dominates the light curve evolution. We also find a clean separation between most simulated TDEs and SNe despite their photometric similarities; a few events are embedded in similar positions as the CSM+Mag and Mag light curves, which may reflect the similarities in the \texttt{MOSFiT} prescriptions for black hole-accretion power (for TDEs) and neutron-star spindown power (for magnetars). These results suggest that our hierarchical SBI model may be able to identify physical analogs to novel populations of transients from LSST light curves.

\section{Conclusions}\label{sec:conclusions}
We have shown that a single hierarchical SBI model can simultaneously identify the energy sources and physical parameters associated with optical time-domain phenomena. This work lays the foundation for scalable population-level studies of transients discovered by existing and upcoming time-domain surveys.  In future work, we will re-train our model with more realistic LSST light curves using LSST Wide-Fast-Deep simulations.\footnote{\url{https://community.lsst.org/t/release-of-v3-4-simulations/8548/7}}. 
We will also investigate the parameter posteriors for partial-phase light curves, which will allow astronomers to rapidly prioritize the most physically-interesting phenomena for real-time follow-up observations. Finally, while Figure~\ref{fig:tsne} shows that the hierarchical model can be used to identify anomalous phenomena, additional work is needed to formalize this approach using density-estimation techniques applied to the learned latent space.

\section{Acknowledgments and Disclosure of Funding}
This work is supported by the National Science Foundation under Cooperative Agreement PHY-2019786 (The NSF AI Institute for Artificial Intelligence and Fundamental Interactions, http://iaifi.org/). The research reported in this paper was done on the Tufts University High Performance Computing Cluster \url{(https://it.tufts.edu/high-performance-computing)}.


\bibliographystyle{unsrtnat} 
\bibliography{references}
\newpage
\appendix

\section{Supernova Multi-Component Priors}
We provide the priors for the parameters of our multi-component physical models below. Where values are not listed, the model defaults are used.
\begin{table}[h]
    \centering
    \begin{tabular}{ccccc}\hline
         Model  &  Parameter & Description & Units & Prior \\ \hline
        -- & $M_{\mathrm{ej}}$ & Ejecta mass & $M_{\odot}$ & $\mathrm{log}\; U(1, 50)$ \\
           & $v_{\mathrm{ej}}$ & Ejecta velocity & $\mathrm{km}\;\mathrm{s}^{-1}$ & $\mathrm{log}\; U(2\times10^{3}, 2\times10^{4})$ \\
           & $L_{\mathrm{dist}}$ & Luminosity distance & Mpc &  $\mathrm{log}\; U(10, 2\times10^{3})$\\
        Ni & $f_{^{56}\mathrm{Ni}}$ & $^{56}$Ni Fraction & -- & $\mathrm{log}\; U(10^{-3}, 10^{-1})$ \\
        Mag & $P_{\mathrm{spin}}$ & Pulsar spin period & ms & $U(0.7, 30)$\\
            & $B_{\mathrm{field}}$ & Magnetic field strength & $10^{14}$~Gauss & $\mathrm{log}\; U(0.1, 15)$ \\
            & $M_{\rm NS}$\footnotemark & Neutron star mass & $M_{\odot}$ & $\mathcal{N}(1.7, 0.2)$\\
        CSM & $n$ & Ejecta density profile index & --  & $U(7, 12)$ \\ 
            & $s$ & CSM density profile index & -- & $U(0.1, 2.0)$\\ 
            & $M_{\mathrm{CSM}}$ & CSM mass & $M_{\odot}$ &  $\mathrm{log}\; U(0.1, 50)$ \\
            & $\rho_0$ & CSM density profile scale & $\mathrm{g}\; \mathrm{cm}^{-3}$ &   $\mathrm{log}\; U(10^{-15}, 10^{-11})$\\ 
        TDE & $T_v$ & Viscous timescale & days & $\mathrm{log}\; U(10^{-3}, 10^2)$ \\ 
         & $M_*$ & Star mass & $M_{\odot}$ & $ U(10^{-1}, 5)$ \\ 
         & $b$ & Scaled impact parameter & -- & $U(0, 2)$ \\ 
         & $M_{BH}$ & Black hole mass & $M_\odot$ & $U(10^6, 10^8)$ \\ 
         & $l_{\mathrm{ph, 0}}$ & Photosphere power-law exponent & -- & $U(0, 2)$ \\
         & $R_{\mathrm{ph},0}$ & Photosphere power-law constant & -- & $\mathrm{log}\; U(10^{-4}, 10^4)$ \\
        \hline
    \end{tabular}
    \caption{Priors for the physical parameters varied for each energy source modeled in this work.}
    \label{tbl:priors}
\end{table}
\footnotetext{We consider the neutron star mass a nuisance parameter, and marginalize over it when sampling from the posterior of models containing a magnetar engine.}

\section{Loss Curves}
\begin{figure}[h]
    \centering
    \includegraphics[width=0.5\textwidth]{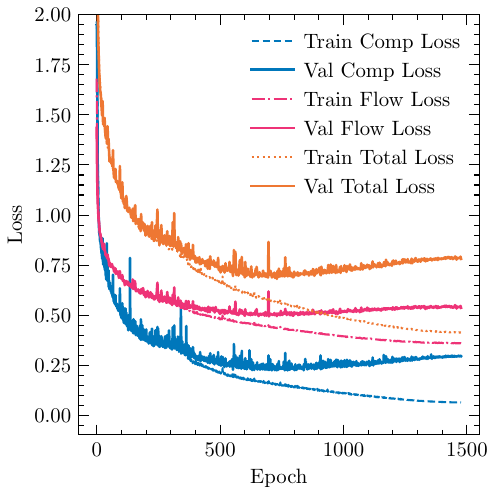}
    \caption{Loss curves for the flow network (pink), categorical network (blue), and combined model (orange) on the training (dashed line) and validation set (solid line).}
    \label{fig:loss_curve}
\end{figure}

We show the resulting training and validation loss curves in Figure~\ref{fig:loss_curve}, split by source component loss (labeled as Comp) and parameter posterior (Flow).

\section{Posterior Coverage Tests}\label{ap:coverage}
\begin{figure}[!htb]
    \centering
    \includegraphics[trim={8cm .1cm .1cm .2cm},clip,width=0.50\textwidth]{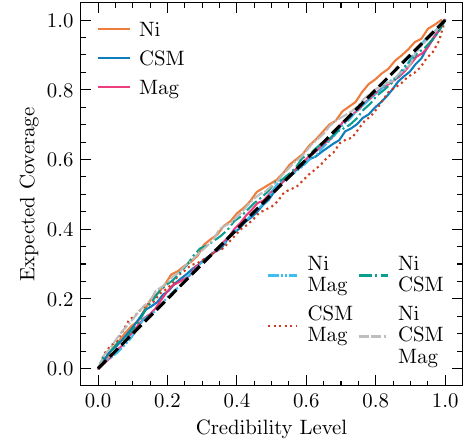}
    \caption{The expected coverage probability compared to the credibility level for the primary components (Ni, CSM, and Mag) and all possible combinations. The black dashed diagonal line indicates perfect calibration.}
    \label{fig:coverage_test}
\end{figure}

The posterior coverage test is shown in Figure~\ref{fig:coverage_test} for each of the different source components, estimated using our test set with \texttt{TARP} 
\citep{lemos2023samplingbasedaccuracytestingposterior}.

\section{Hyperparameter Tuning}\label{ap:hyperparameters}
We provide the hyperparameters used in our model, obtained through a Bayesian optimization sweep implemented with \texttt{wandb} \citep{wandb}.
\begin{table}[!htb]
\centering
\caption{Hyperparameter sweep configuration and selected optimal hyperparameter values.}
\label{tab:bayes_sweep}
\begin{tabular}{lll}
\hline
Parameter & Search Values / Range & Selected Value \\
\hline
\multicolumn{3}{l}{\textit{Summarizer hyperparameters}} \\
\hspace{1em}hidden dimension & [32, 64, 128, 256] & 64 \\
\hspace{1em}latent dimension & [16, 32, 64, 128] & 64 \\
\hspace{1em}transformer heads & [2, 4] & 2 \\
\hspace{1em}transformer depth & [2, 4, 8] & 4 \\
\hspace{1em}dropout rate & [0.0, 0.01, 0.02] & 0.0 \\
\\[-0.5em]
\multicolumn{3}{l}{\textit{Velocity network hyperparameters}} \\
\hspace{1em}hidden dimension & [64, 128, 256] & 64 \\
\hspace{1em}latent dimension & [32, 64, 128] & 128 \\
\hspace{1em}transformer heads & [2, 4] & 4 \\
\hspace{1em}transformer depth & [2, 4, 8] & 8 \\
\\[-0.5em]
\multicolumn{3}{l}{\textit{Categorical network hyperparameters}} \\
\hspace{1em}hidden dimension & [32, 64, 128] & 128 \\
\hspace{1em}number of layers & [2, 4] & 4 \\
\\[-0.5em]
\multicolumn{3}{l}{\textit{Joint training hyperparameters}} \\
\hspace{1em}classification weight & U(0.5, 1.2) & 0.8264 \\
\\[-0.5em]
\multicolumn{3}{l}{\textit{Optimization}} \\
\hspace{1em}learning rate & [1e--4, 3e--4] & 1e--4 \\
\hspace{1em}weight decay & [0.0, 1e--5, 1e--4] & 0.0 \\
\hspace{1em}batch size & [256] & 256 \\
\\[-0.5em]

\hline
\end{tabular}
\end{table}

\end{document}